\documentclass[3p,super,numbers,sort&compress,preprint,10pt]{elsarticle}
\makeatletter\if@twocolumn\PassOptionsToPackage{switch}{lineno}\else\fi\makeatother

\usepackage{tabulary,xcolor}
\usepackage{amsfonts,amsmath,amssymb}
\usepackage[T1]{fontenc}
\makeatletter
\let\save@ps@pprintTitle\ps@pprintTitle
\def\ps@pprintTitle{\save@ps@pprintTitle\gdef\@oddfoot{\footnotesize\itshape \null\hfill\today}}
\def\hlinewd#1{%
  \noalign{\ifnum0=`}\fi\hrule \@height #1%
  \futurelet\reserved@a\@xhline}

\def\tblbottomrule{\noalign{\vspace*{6pt}}\hline\noalign{\vspace*{2pt}}}
\def\tblmidrule{\noalign{\vspace*{6pt}}\hline\noalign{\vspace*{2pt}}}
\AtBeginDocument{\ifNAT@numbers \biboptions{sort&compress}\fi}
\makeatother

\usepackage{caption}
\usepackage{times}

\def\NormalBaseline{\def\baselinestretch{1.1}}

\makeatletter

\def\title#1{\gdef\@title{\fontsize{12pt}{14pt}\selectfont {\bfseries #1}}}
\makeatother

\usepackage[noindentafter]{titlesec}

\titleformat{\section}[hang]{\NormalBaseline\filright\fontsize{10}{12}\boldmath\bfseries\selectfont}
{\fontsize{10}{12}\selectfont\thesection}
{10pt}
{\noindent}
[]
\titleformat{\subsection}[hang]{\NormalBaseline\filright\fontsize{10}{12}\boldmath\bfseries\selectfont}
{\fontsize{10}{12}\selectfont \thesubsection}
{10pt}
{}
[]
\titleformat{\subsubsection}[hang]{\NormalBaseline\filright\fontsize{10}{11}\itshape\selectfont}
{\fontsize{10}{11}\itshape\selectfont\thesubsubsection}
{10pt}
{}
[]
\titleformat{\paragraph}[runin]{\NormalBaseline\boldmath\bfseries}
{\theparagraph}
{10pt}
{}
[]
\titleformat{\subparagraph}[runin]{\NormalBaseline\boldmath\bfseries\itshape}
{\thesubparagraph}
{10pt}
{}
[]

\usepackage{fancyhdr}

\fancypagestyle{headings}{\fancyhf{}\fancyfoot[R]{\thepage}}

\fancypagestyle{pprintTitle}{\fancyhf{}\fancyfoot[R]{\thepage}}
\usepackage{ifluatex}
\ifluatex
\usepackage{fontspec}
\defaultfontfeatures{Ligatures=TeX}
\usepackage[]{unicode-math}
\unimathsetup{math-style=TeX}
\else 
\usepackage[utf8]{inputenc}
\fi 
\ifluatex\else\usepackage{stmaryrd}\fi

\usepackage{url,multirow,morefloats,floatflt,cancel,tfrupee}
\makeatletter

\AtBeginDocument{\@ifpackageloaded{textcomp}{}{\usepackage{textcomp}}}
\makeatother
\usepackage{colortbl}
\usepackage{xcolor}
\usepackage{pifont}
\usepackage[nointegrals]{wasysym}
\makeatletter

\def\mcWidth#1{\csname TY@F#1\endcsname+\tabcolsep}

\def\cAlignHack{\rightskip\@flushglue\leftskip\@flushglue\parindent\z@\parfillskip\z@skip}
\def\rAlignHack{\rightskip\z@skip\leftskip\@flushglue \parindent\z@\parfillskip\z@skip}

\@ifundefined{etal}{}{}

\usepackage{ifxetex}
\ifxetex\else\if@twocolumn\@ifpackageloaded{stfloats}{}{\usepackage{dblfloatfix}}\fi\fi

\AtBeginDocument{
\expandafter\ifx\csname eqalign\endcsname\relax
\def\eqalign#1{\null\vcenter{\def\\{\cr}\openup\jot\m@th
  \ialign{\strut$\displaystyle{##}$\hfil&$\displaystyle{{}##}$\hfil
      \crcr#1\crcr}}\,}
\fi
}

\AtBeginDocument{%
  \@ifpackageloaded{endfloat}%
   {\renewcommand\efloat@iwrite[1]{\immediate\expandafter\protected@write\csname efloat@post#1\endcsname{}}}{\newif\ifefloat@tables}%
}%

\def\BreakURLText#1{\@tfor\brk@tempa:=#1\do{\brk@tempa\hskip0pt}}
\let\lt=<
\let\gt=>
\def\processVert{\ifmmode|\else\textbar\fi}

\@ifundefined{subparagraph}{
\def\subparagraph{\@startsection{paragraph}{5}{2\parindent}{0ex plus 0.1ex minus 0.1ex}%
{0ex}{\normalfont\small\itshape}}%
}{}

\newcommand\role[1]{\unskip}
\newcommand\aucollab[1]{\unskip}
  
\@ifundefined{tsGraphicsScaleX}{\gdef\tsGraphicsScaleX{1}}{}
\@ifundefined{tsGraphicsScaleY}{\gdef\tsGraphicsScaleY{.9}}{}
\def\checkGraphicsWidth{\ifdim\Gin@nat@width>\linewidth
	\tsGraphicsScaleX\linewidth\else\Gin@nat@width\fi}

\def\checkGraphicsHeight{\ifdim\Gin@nat@height>.9\textheight
	\tsGraphicsScaleY\textheight\else\Gin@nat@height\fi}

\def\fixFloatSize#1{}
\let\ts@includegraphics\includegraphics

\def\inlinegraphic[#1]#2{{\edef\@tempa{#1}\edef\baseline@shift{\ifx\@tempa\@empty0\else#1\fi}\edef\tempZ{\the\numexpr(\numexpr(\baseline@shift*\f@size/100))}\protect\raisebox{\tempZ pt}{\ts@includegraphics{#2}}}}

\AtBeginDocument{\def\includegraphics{\@ifnextchar[{\ts@includegraphics}{\ts@includegraphics[width=\checkGraphicsWidth,height=\checkGraphicsHeight,keepaspectratio]}}}

\DeclareMathAlphabet{\mathpzc}{OT1}{pzc}{m}{it}

\def\URL#1#2{\@ifundefined{href}{#2}{\href{#1}{#2}}}

\def\UrlOrds{\do\*\do\-\do\~\do\'\do\"\do\-}%
\g@addto@macro{\UrlBreaks}{\UrlOrds}

\edef\fntEncoding{\f@encoding}

\makeatother

\newif\ifmultipleabstract\multipleabstractfalse%
\usepackage{longtable}

\makeatletter
\AtBeginDocument{\@ifpackageloaded{rotating}{\PassOptionsToPackage{figuresright}{rotating}}{\usepackage[figuresright]{rotating}}\setlength{\rotFPtop}{0pt plus 1fil}\setlength{\rotFPbot}{0pt plus 1fil}}
\makeatother

\makeatletter
 \AtBeginDocument{%
  \@ifpackagewith{endfloat}{figuresonly}
  {\DeclareDelayedFloatFlavor{sidewaysfigure}{figure}}
  {\@ifpackagewith{endfloat}{tablesonly}{\DeclareDelayedFloatFlavor{sidewaystable}{table}\DeclareDelayedFloatFlavor{longtable}{table}\DeclareDelayedFloatFlavor{landscape}{table}}
  {\@ifpackageloaded{endfloat}{\DeclareDelayedFloatFlavor{sidewaysfigure}{figure}\DeclareDelayedFloatFlavor{sidewaystable}{table}\DeclareDelayedFloatFlavor{longtable}{table}\DeclareDelayedFloatFlavor{landscape}{table}}{}}
  }
  }
\makeatother

\usepackage{pdflscape}

\begin{document}

\begin{frontmatter}

\title{
    \textbf{Perceptions and attitudes of Children and Young People to Artificial Intelligence in Medicine }    
}
    
\author[a7e47bc15c08c,af7ddc1d0c509]{Sheena Visram\corref{contrib-eb6992ff993e44acb685bbc75af56b70}}
\ead{s.visram@ucl.ac.uk}\cortext[contrib-eb6992ff993e44acb685bbc75af56b70]{Corresponding author.}
\author[a1ca8ebd71c07]{Deirdre Leyden}
\author[a1ca8ebd71c07]{Oceiah Annesley}
\author[a1ca8ebd71c07]{Dauda Bappa}
\author[af7ddc1d0c509]{Neil J Sebire\corref{contrib-96cf5f7efbee4b36b387d99ac8afaddb}}
\ead{neil.sebire@gosh.nhs.uk}\cortext[contrib-96cf5f7efbee4b36b387d99ac8afaddb]{Corresponding author.}
    
\address[a7e47bc15c08c]{Department of Computer Science |~UCLIC\unskip, 
    University College London\unskip, London\unskip, United Kingdom}
  	
\address[af7ddc1d0c509]{DRIVE centre\unskip, 
    Great Ormond Street Hospital for Children \unskip, London\unskip, United Kingdom}
  	
\address[a1ca8ebd71c07]{Young Persons Advisory Group (YPAG)\unskip, 
    Great Ormond Street Hospital for Children\unskip, London\unskip, United Kingdom}

\begin{abstract}
Introduction: There is increasing interest in Artificial Intelligence (AI) and its application to medicine. Perceptions of AI are less well-known, notably amongst children and young people. This exploratory patient and public engagement (PPEI) workshop investigates attitudes towards AI and its future applications in medicine and healthcare from the perspective of children and young people with lived experiences at a specialised paediatric hospital using practical design scenarios. \textbf{Method: }Members of Great Ormond Street Hospital for Children's (GOSH), Young Persons Advisory Group for research (YPAG) contributed to a one-hour AI workshop to ascertain potential opportunities, apprehensions, and priorities. Quantitative polling using a series of nine AI-related design scenarios were scored voluntarily and anonymously on a 10-point Likert scale and mechanisms for effectively engaging with patients and families on the potential for AI were discussed. \textbf{Results: }21 GOSH YPAG members (aged 10-21 years) participated. Human-centeredness, governance and trust emerged as major themes, with empathy and safety considered as important when introducing AI to healthcare.  Of the scenarios presented, sensor technology to reduce overcrowding (M 7.4, SD 2.7), cleaning robots (M 7.9, SD 2.4), virtual reality visits (M 6.5, SD 2.8) and 3D printed organs (M 6.2, SD 3.5) were the most acceptable, whilst AI-powered nurses the least acceptable (M 2.4, SD 2.3). Educational workshops with practical examples using AI to help, but not replace, humans were suggested to address issues, build trust and effectively communicate about AI. \textbf{Conclusion: }Whilst policy guidelines acknowledge the need to include children and young people to develop AI, this ignores infrastructure needs to encourage digital cooperation. For AI in medicine and healthcare this requires an enabling environment for human-centred AI involving children and young people with lived experiences of healthcare\textbf{. } This PPEI workshop is an important mechanism to shape future research questions. Future research should focus on building consensus on enablers for an intelligent healthcare system designed for the next generation, which fundamentally, allows co-creation.
\end{abstract}
\end{frontmatter}
    
\section{Introduction}
There is growing interest in the application of Artificial Intelligence (AI) to medicine. Initially described as exotic, expensive, and not of benefit to ordinary people\ensuremath{^{1}}, global interest within the field has increased exponentially\ensuremath{^{2}}. High-quality reviews of AI in healthcare have addressed its use, value, and trustworthiness \ensuremath{^{3\textit{\ensuremath{-}}6}}. In children's healthcare, parents ask for openness during AI development, and ask  that technical experts consider shared decision making, the human element of care and social justice as part of the development process\ensuremath{^{7}}. However, whilst views of Children and Young People (CYP) can shape healthcare provision \ensuremath{^{8\textit{\ensuremath{-}}11}}, few policy recommendations reflect their views and beliefs\ensuremath{^{12\textit{\ensuremath{-}}14}}. This is particularly the case for CYP with tacit healthcare knowledge.

Great Ormond Street Hospital for children (GOSH) is the largest paediatric centre in the UK and an international centre of excellence for many clinical specialties. As part of the hospital, the Digital Research, Informatics, and Virtual Environments (DRIVE) unit aims to accelerates research and deployment of new technology including working with patients and families to optimise technologies such as AI. The Young Persons Advisory Group (YPAG) is a patient and public involvement group embedded at the hospital comprising CYP who are interested in improving health by advising on research, and forms part of a national network (Generation R).

Using a workshop entitled AI\&me, we address a current deficit in AI healthcare policy and practice, by exploring the perspective of CYP with lived experiences of healthcare including establishing priorities of GOSH YPAG in an exploratory PPEI design workshop on Healthcare AI.
    
\section{Method}
A single design workshop examined perceptions and attitudes of CYP on AI applications in medicine and healthcare. Findings were reported using the COREQ 32-point checklist for focus group reflexivity, design and analysis (included as a supplement)\ensuremath{^{ 18 }}

\subsection{Sampling}Members of GOSH YPAG contributed to an exploratory engagement workshop run virtually, lasting one hour to explore their perceptions of AI in medicine and healthcare. They rated levels of comfort with AI-related design scenarios and discussed mechanisms to effectively engage with patients and families on AI's future potential.

\subsection{Design}The virtual workshop opened with a short broad discussion about AI, after which nine design scenarios were presented, including: Virtual Reality visits to hospitals, cleaning robots, talking robots, chatbots to diagnose disease, self-driving vehicles, AI-powered nurses, 3D printed hearts and sensor technology to reduce overcrowding.  These were developed from a recent survey of 2000 parents\ensuremath{^{15 }}and predominantly focused on healthcare applications of technologies intended to delight, inform, predict, automate or diagnose/treat.

\subsection{Data collection}Quantitative polling of scenarios was undertaken anonymously using a 10-point Likert-scale. To collect comments, a virtual chat function and an agile, Audience Response System (Mentimeter AB, Stockholm, Stockholms Lan, Sweden) were used since these are effective for encouraging participation in virtual learning environments\ensuremath{^{16\textit{,} 17}} . Comments were collected verbatim.

\subsection{Data analysis}Inductive qualitative content analysis identified concepts and emergent themes using NVivo  for Windows v.1.4.1 (QSR Inter- national, Melbourne, Australia) to shape future research questions, as there is limited existing qualitative research regarding perceptions of AI in medicine and healthcare amongst CYP with lived experiences. This involved data familiarisation, immersion and iterative identification of codes, concepts, phrases and language. Open codes were collated under emerging themes and findings supported by verbatim quotes.
    
\section{Results}
21 YPAG members (aged 10- 21 years) participated, generating 128 unique comments across platforms. 

The language used by participants comprised words that described how AI made them feel (58 generalised occurrences that included affect, care, compassion, consider, experience and fear), AI was commonly referred to as a `robot' (18 incidents) and `creepy' on six occasions. Patients were commonly mentioned (18 occurrences) and generalised words relating to comfort (assure and reassure) were used 26 times. The comments were conversational, but several comments were structured as questions (n=28, 22\%) suggesting interest to understand more about AI (Figure 1b).

\subsection{Design scenarios}Of the nine design scenarios presented, sensor technology to reduce overcrowding (M 7.4, SD 2.7), cleaning robots (M 7.9, SD 2.4), virtual reality visits (M 6.5, SD 2.8) and 3D printed organs (M 6.2, SD 3.5) were the most accepted scenarios, whilst AI-powered nurses the least (M 2.4, SD 2.3; Figure 1c).

\subsection{Emerging themes}Three themes emerged from the exploratory engagement workshop: governance, human centredness and trust (Figure 1a).

\subsubsection{Governance}Safety and benefits formed the basis of a number of early inquiries about AI. There was an interest that access to AI-enabled technologies was fair and available to all. Ensuring safety, security risks, and reliability was of particular interest, one participant asking:

``What safety measures are in place?'', another:

`What happens if the robot makes(s) a mistake or the software breaks down?'', expanding to ask: ``Would the robot get the benefit of the doubt?''

More broadly, on ethical use of AI, one participant asked:

``How do you stop people abusing the system?''

As members of YPAG at a specialist paediatric hospital, a number of questions were raised about the role of AI for rare diseases, and potential benefits to challenges faced in healthcare, one participant asking:

 ``Will it speed up waiting times in A\&E'' and on effectiveness, one participant asked:

``If a rare disease occurs, how will the robot know what to do as there is no specific treatment'', another:

``How do you train AI if someone develops a new illness'' and:

``Is an online chat bot actually more beneficial to patients?''

\subsubsection{Human Centredness}The role of human-centred care in healthcare was another emergent theme with empathy, agency and power dynamics considered important. It was thought that AI would not take emotions into account and this could have an impact on treatment, especially where mental health and wellbeing are considered. One participant asked:

``How do you teach AI to be empathetic and understand pain?'' another:

``How would bad news be broken to patients?''

Agency and control over the use of AI was a pertinent topic, one participant reflecting:

``I like the idea of AI looking at scans and in surgery, but definitely not for decision making or patient interaction'', another asking:

``Would AI make the decision or be the advisor to the doctor?'' and another:

``Would doctors be able to overrule AI if they're not happy with the decision/ course of action?'' 

Replacing humans was commonly associated to the impact on jobs, one participant expressing: ``I don't like the idea of robots taking jobs'' and another asked:

``What will happen to the doctors who are working now?'' expanding to: ``will their jobs get replaced?''.

Another participant reflected on the potential impact on skillset and disparities between countries using AI and others that do not, asking:

``Will doctors need to be less qualified if the use of AI is normalised?''

A popular remark anticipated the role of AI as supportive rather than to replace healthcare staff, one participant stating: ``I will find it ok as long as it is just helping and does not replace humans'',

\subsubsection{Trust}The influence of movies, games and science fiction on perceptions of AI was a popular topic. Opinions on AI amongst children and young people are influenced by pop culture and science fiction, which often depict robots as evil, one participant reflecting:

``I think we watch too many sci-fi movies'', another ``that's why I'm scared of robots'' This led to comments about creepiness, one participant stating:

``AI is creepy if it acts like a human''

Educational workshops with reassurance, practical examples that use AI to help, but not replace humans were suggested to address common worries, build trust and to effectively communicate about AI. To cultivate trust, it was recommended that healthcare staff are transparent about its use, with clear explanations and examples of its use in everyday life (Figure 1d), one participant recommended:

``Being transparent when you are already using it e.g., when AI is used in conjunction with surgeons'' with ``success stories \& when things go wrong \& how it was resolved''.

Ethical considerations about who would make decisions and what might happen should something go wrong were considered. One participant stating:

``Make sure you address common worries instead of avoiding them when explaining AI''.

Overall, participants were interested to engage on further discussions about AI, and a generational gap was identified, that considers young people more open to and comfortable with AI in general.

``YPAG members are keen to be involved, for our perspective and ideas, especially as AI is our future''.
    
\section{Discussion}
The findings of this exploratory workshop, intended to infrom furture research have demonstrated that CYP are open-minded to using AI in medicine and healthcare and believe that this technology will change everyday life in fundamental ways but find it difficult to articulate their views on how AI should be developed. This is partly due to the breadth of applications and their impacts\ensuremath{^{19}}. CYP need to be educated about AI and encouraged to participate in its development including making AI explainable to children and young people by including them in AI policy development cycles\ensuremath{^{12}}. 

Outside of healthcare, UNICEF recommends nine requirements for child-centred AI, including inclusion, safety, privacy, transparency, and the need to create an enabling environment to discover whether AI systems are designed for children and potential impacts\ensuremath{^{12}}. Whilst participatory research is described as a key element, such policy guidelines are not accompanied by practical recommendations to enable such digital cooperation.

This is the first exploration through a virtual, group-based workshop to engage with CYP with lived experiences of healthcare regarding perceptions of AI in medicine and healthcare. We demonstrate that by creating an open and enabling environment and using design scenarios to discuss potential applications, YPAG members were keen to participate, share opinions, outline concerns, and further develop their own understanding of AI.

By including and involving CYP in this space, we can optimise AI to enhance future experiences of care\ensuremath{^{20}}. To achieve this shared aspiration requires collaboration, and where there are areas of disagreement or uncertainty, these need to be clearly identified. This involves creating an enabling environment for CYP-centred AI and involving CYP with lived experiences of healthcare in the process in ways that engage, inspire and empower\ensuremath{^{13,21}}.

\subsection{Limitations}The preliminary findings reported reflect a PPEI workshop intended to inform future research and spur deliberations on this topic. Whilst content analysis represents an appropriate analytic approach which is unobtrusive, nonreactive and time-efficient when compared to methods such as ethnography, this workshop in its design was limited to the breadth of specific potential AI applications discussed, and by the depth of discussion achieved during a virtual one-hour group session. Data saturation was not intended to be achieved, rather emergent themes identified to shape future research. Whilst the design of this exploratory workshop allowed for rapid, and well attended virtual participation, future research approaches might also include supplementary in-depth interviews and consensus building methods.
    
\section{Conclusion}
CYP want to be included in the development of AI in medicine and healthcare. Whilst policy guidelines acknowledge the need to include CYP this ignores the infrastructure required to support ongoing digital cooperation. For AI in medicine, this requires an enabling environment for human-centred AI that involves CYP with lived experiences of healthcare and healthcare/AI professionals. With publication of the recent UK National Strategy for AI, future research should explore the ways CYP can participate in shaping an intelligent, empathetic, and inclusive healthcare system of tomorrow and in the application and development of AI in healthcare.

\section{References}
\textbf{\space }

  \begin{enumerate}
  \item \relax Coles LS. The application of artificial intelligence to medicine. Futures. 1977 Aug; 9(4):315{\textendash}23.
  \item \relax Chang AC. Intelligence-Based Medicine: Artificial Intelligence and Human Cognition in Clinical Medicine and Healthcare. Academic Press; 2020. 549 p.
  \item \relax Hamet P, Tremblay J. Artificial intelligence in medicine. Metabolism. 2017 Apr 1; 69:S36{\textendash}40.
  \item \relax Szolovits P. Artificial Intelligence In Medicine. Routledge; 2019. 255 p.
  \item \relax Davendralingam N, Sebire NJ, Arthurs OJ, Shelmerdine SC. Artificial intelligence in paediatric radiology: Future opportunities. Br J Radiol. 2020 Sep 17; 94(1117):20200975.
  \item \relax Liang H, Tsui BY, Ni H, Valentim CCS, Baxter SL, Liu G, et al. Evaluation and accurate diagnoses of pediatric diseases using artificial intelligence. Nat Med. 2019 Mar; 25(3):433{\textendash}8.
  \item \relax Sisk BA, Antes AL, Burrous S, DuBois JM. Parental Attitudes toward Artificial Intelligence-Driven Precision Medicine Technologies in Pediatric Healthcare. Children [Internet]. 2020 Sep 20 [cited 2021 Apr 29]; 7(9). Available from: 
  \item \relax Weil 
  \item \relax WHO |~Making health services adolescent friendly [Internet]. WHO. World Health Organization; [cited 2021 Apr 29]. Available from: 
  \item \relax Hargreaves DS, Lemer C, Ewing C, Cornish J, Baker T, Toma K, et al. Measuring and improving the quality of NHS care for children and young people. Arch Dis Child. 2019 Jul; 104(7):618{\textendash}21.
  \item \relax Well 
  \item \relax UNICEF policy guidance on AI for children [Internet]. The Commonwealth. [cited 2021 Jul 18]. Available from: 
  \item \relax Adolescent perspectives on artificial intelligence [Internet]. [cited 2021 Apr 29]. Available from: \BreakURLText{https://www.un} icef.org/globalinsight/stories/adolescent-perspectives-artificial-intelligence
  \item \relax Artificial Intelligence for Children: Beijing Principles [Internet]. [cited 2021 Jun 10]. Available from: \BreakURLText{https://ww} w.baai.ac.cn/ai-for-children.html
  \item \relax Generation AI 2020: Health, Wellness and Technology in a Post-COVID World [Internet]. IEEE Transmitter. 2020 [cited 2021 Jun 10]. Available from: \BreakURLText{https://transmitter.ieee.org/generation-ai-2020/}
  \item \relax Mayhew E, Davies M, Millmore A, Thompson L, Bizama AP. The impact of audience response platform Men- timeter on the student and staff learning experience. Res Learn Technol [Internet]. 2020 Oct 30 [cited 2021 Jul 7]; 28. Available from: \BreakURLText{https://journal.alt.ac.uk/index.php/rlt/article/view/2397}
  \item \relax Little C. Mentimeter Smartphone Student Response System: A class above clickers. Compass J Learn Teach [Internet]. 2016 Nov 8 [cited 2021 Jul 16]; 9(13). Available from: \BreakURLText{https://journals.gre.ac.uk/index.php/compass/articl} e/view/328
  \item \relax Tong, Sainsbury P, Craig J. Consolidated criteria for reporting qualitative research (COREQ): a 32-item check- list for interviews and focus groups. Int J Qual Health Care. 2007 Dec 1; 19(6):349{\textendash}57.
  \item \relax Cameron D, Maguire K. Public views of machine learning: Digital Natives. Available from: \BreakURLText{https://royalsociety.org/-/media/policy/projects/machine-learning/digital-natives-16-10-2017.pdf}
  \item \relax Hargreaves DS, Sizmur S, Pitchforth J, Tallett A, Toomey SL, Hopwood B, et al. Children and young people's versus parents' responses in an English national inpatient survey. Arch Dis Child. 2018 May; 103(5):486{\textendash}91.
  \item \relax Freire, K \& Sangiorgi, D. Service design and healthcare innovation: from consumption, to co-production to co-creation. 2010. Paper presented at Nordic Service Design Conference, Linkoping, Sweden
  \end{enumerate}

\section{Acknowledgments}
We thank the 21 members of GOSH YPAG for their engagement in this exploratory workshop. This work is supported by the NIHR GOSH BRC. 
    
\section{Contributions}
SV, NJS conceived the project. DL, DB were involved with arranging the YPAG meeting and inviting interested YPAG members to engage. SV, NJS led the facilitation of the YPAG session and data collection. DB collected independent observations to triangulate findings. SV conducted the analysis and reported on findings in compliance with the COREQ 32-point checklist, and creating Figure 1. OA represented the viewpoint of a YPAG member to validate emerging themes, add reference to co-creation, and review the manuscript for layman language and style. All authors contributed to the validation process, checking for accurate representation of findings, completeness and provided revisions to early drafts of the manuscript.
    
\section{Funding statement}
SV acknowledges doctoral funding from Great Ormond Street Hospital  for Children charities.

\begin{landscape}
\makeatletter\@twocolumnfalse\makeatother
\begingroup
\makeatletter\if@twocolumn\@ifundefined{theposttbl}{\gdef\TwoColDocument{true}\onecolumn\onecolumn}{}\fi\makeatother \setlength\LTcapwidth{\textheight}
\begin{longtable}{p{\dimexpr.33\linewidth-2\tabcolsep}p{\dimexpr.33\linewidth-2\tabcolsep}p{\dimexpr.34\linewidth-2\tabcolsep}}
\caption{{COREQ 32-point checklist for reporting on focus groups\ensuremath{^{18}}} }
\label{table-wrap-05280b11bf924f68b0e1546a68299b94}
\fontsize{8pt}{10pt}\selectfont \\\endfirsthead \hline \noalign{\vskip3pt} \noalign{\textit{Table \thetable\ continued}} \noalign{\vskip3pt} 
\hline  &  & \\
\tblmidrule \endhead \hline \noalign{\vskip3pt} \noalign{\textit{\hfill Continued on next page}} \noalign{\vskip3pt} \endfoot \endlastfoot 
\hline  &  & \\
\tblmidrule 
\textbf{No} &
  \textbf{Item} &
  \textbf{Description}\\
\textbf{Domain 1: Research team and reflexivity} &
   &
  \\
 Personal Characteristics  &
   &
  \\
 1.  &
   Interviewer/facilitator  &
   Facilitators (SV, NJS) \\
 2.  &
   Credentials  &
   Undertaking PhD in Computer Science/ Human Computer Interaction (SV), Professor of Pathology and Chief Research and Informatics Officer (CRIO) (NJS) \\
 3.  &
   Occupation  &
   PhD candidate and visiting researcher to GOSH (SV), Professor of Pathology, CRIO and Director of GOSH DRIVE (NJS)\\
 4.  &
   Gender  &
   Female (SV), Male (NJS) \\
 5.  &
   Experience and training  &
  Experienced - Undertaking PhD in Computer Science/ Human Computer Interaction (SV), Professor of Pathology and Chief Research and Informatics Officer (CRIO) (NJS) \\
 Relationship with participants  &
   &
  \\
 6.  &
   Relationship established  &
  Prototypes of new technologies are presented to YPAG in their role to provide feedback on research and so the participants were familiar with who the facilitators were.\\
 7.  &
   Participant knowledge of the interviewer  &
  Participants knew the credentials of the facilitators and that SV is conducting research on human-centred aspects of technology adoption. \\
 8.  &
   Interviewer characteristics  &
  The facilitators are interested in new technologies. This is an exploratory workshop and did not intend to prove or disprove a hypothesis, but instead garner perceptions of children and young people to shape future research.\\
\textbf{Domain 2:  Design} &
   &
  \\
 Theoretical framework  &
   &
  \\
 9.  &
   Methodological orientation and Theory  &
  Descriptive statistics of quantitative polling of design scenarios involving AI, and Content analysis of 128 short comments or micronarratives. \\
 Participant selection  &
   &
  \\
 10.  &
   Sampling  &
  Purposive selection as members of YPAG \\
 11.  &
   Method of approach  &
  Email invitation by Patient and Public Involvement Research Lead, NIHR GOSH Biomedical Research Centre \\
 12.  &
   Sample size  &
   21 \\
 13.  &
   Non-participation  &
   0 {\textendash} participation was voluntary for this engagement session \\
 Setting  &
   &
  \\
 14.  &
   Setting of data collection  &
  Virtually via a video conferencing platform with comments posted anonymously to an audience response system and in chat functions.\\
 15.  &
   Presence of non-participants  &
  Patient and Public Involvement Research Lead, NIHR GOSH Biomedical Research Centre (1) , Biomedical Research Centre staff (1), Biomedical Research Centre staff as a note taker for field notes (1)\\
 16.  &
   Description of sample  &
  Children and young people ages 10-21 years of age, typically with lived experiences of healthcare and members of YPAG, an advisory group that feeds back on research. \\
 Data collection  &
   &
  \\
 17.  &
   Interview guide  &
  The session was introduced with an agenda two months prior to the session and the questions were circulated to the team at GOSH DRIVE \\
 18.  &
   Repeat interviews  &
   No \\
 19.  &
   Audio/visual recording  &
   No \\
 20.  &
   Field notes  &
  One independent observer took field notes during the session which were used to cross reference codes and emerging themes from the data \\
 21.  &
   Duration  &
   One hour \\
 22.  &
   Data saturation  &
  This was an exploratory workshop and so data saturation was not anticipated \\
 23.  &
   Transcripts returned  &
  Comments made in the chat function were made available to participants, and comments made anonymously on the audience response systems displayed as a rolling grid in real time during the session. \\
\textbf{Domain 3: analysis and findings} &
   &
  \\
 Data analysis  &
   &
  \\
 24.  &
   Number of data coders  &
   1 (SV) \\
 25.  &
   Description of the coding tree  &
  Yes, for each emerging theme (human centredness, governance and trust)   Coding tree on NVivo (Figure 2)\\
 26.  &
   Derivation of themes  &
  Derived inductively from the data \\
 27.  &
   Software  &
  NVivo for Windows v.1.4.1 (QSR International, Melbourne, Australia)\\
 28.  &
   Participant checking  &
  One participant was invited to cross check emerging codes and themes for accuracy and co-author the paper that presented preliminary findings\\
 Reporting  &
   &
  \\
 29.  &
   Quotations presented  &
   Yes, quotations were presented, the participants were not identified as comments were made anonymously. This was intended to encourage open and honest discussions \\
 30.  &
   Data and findings consistent  &
   Yes \\
 31.  &
   Clarity of major themes  &
  Yes, for each emerging theme (human centredness, governance and trust) \\
 32.  &
   Clarity of minor themes  &
  Yes, for divergence in opinions and open coding \\
\tblbottomrule 
\end{longtable}
\endgroup
\makeatletter\@ifundefined{TwoColDocument}{}{\twocolumn}\makeatother 
\end{landscape}

\bibliographystyle{vancouver}

\bibliography{\jobname}

\end{document}